\documentclass{aa}  

\usepackage{graphicx}
\usepackage{txfonts}
\usepackage{verbatim}
\usepackage{natbib}
\usepackage{amsmath}

\def\kms{km\,s$^{-1}$}

\def\msun{M$_{\odot}$}
\def\msol{M$_{\odot}$}
\def\rsol{R$_{\odot}$}
\def\rsun{R$_{\odot}$}

\def\l{$\lambda$}

\def\ha{H\,{\sc i}}

\def\hea{He\,{\sc i}}
\def\heb{He\,{\sc ii}}

\def\cd{C\,{\sc iv}}

\def\fw{\textsc{fastwind} }
\def\phoebe{\textsc{phoebe} }
\def\spamms{\textsc{spamms} }

\begin{document}

   \title{Spectroscopic patch model for massive stars using \textsc{phoebe ii} and \fw}


   \author{Michael Abdul-Masih\inst{1},
          Hugues Sana\inst{1},
          Kyle E. Conroy\inst{2},
          Jon Sundqvist\inst{1},
          Andrej Pr\v sa\inst{2},
          Angela Kochoska\inst{2},
          Joachim Puls\inst{3}
          }

   \institute{Institute of Astrophysics, KULeuven, Celestijnenlaan 200 D, 3001 Leuven, Belgium\\
              \email{michael.abdulmasih@kuleuven.be}
         \and
             Dept. of Astrophysics and Planetary Science, Villanova University, 800 Lancaster Ave, Villanova, 19085, USA
         \and
             LMU München, Universitätssternwarte, Scheinerstr. 1, 81679 München, Germany
             }

   \date{Received December 17, 2019; accepted February 18, 2020}

 
  \abstract
   {Massive stars play an important role in the mechanical and chemical evolution of galaxies. Understanding the internal processes of these stars is vital to our understanding of their evolution and eventual end products.  Deformations from spherical geometry are common for massive stars; however, the tools that are currently available for the study of these systems are almost exclusively one-dimensional.}
   {We present a new spectroscopic analysis tool tailored for massive stars that deviate from spherical symmetry. This code (entitled \textsc{spamms}) is a spectroscopic patch model that takes the three-dimensional surface geometry of the system into account to produce spectral profiles at given phases and orientations\thanks{github.com/MichaelAbdul-Masih/SPAMMS}.}
   {In using the Wilson-Devinney-like code \textsc{phoebe} in combination with the nonlocal thermodynamic equilibrium (NLTE) radiative transfer code \textsc{fastwind}, we created a three-dimensional mesh that represents the surface geometry of our system and we assigned \fw emergent intensity line profiles to each mesh triangle, which take the local parameters such as temperature, surface gravity, and radius into account.  These line profiles were then integrated across the visible surface, where their flux contribution and radial velocity are taken into account, thus returning a final line profile for the visible surface of the system at a given phase.}
   {We demonstrate that \spamms can accurately reproduce the morphology of observed spectral line profiles for overcontact systems.  Additionally, we show how line profiles of rapidly-rotating single stars differ when taking rotational distortion into account, and the effects that these can have on the determined parameters.  Finally, we demonstrate the code's ability to reproduce the Rossiter-Mclaughlin and Struve-Sahade effects.}
  {}

   \keywords{binaries: close --
             binaries: eclipsing -- 
   			 binaries: spectroscopic -- 
   			 stars: massive -- 
   			 stars: rotation -- 
   			 techniques: spectroscopic
               }
	\titlerunning{SPAMMS}
	\authorrunning{M. Abdul-Masih et al.}
   \maketitle
%

\section{Introduction}

Due to their strong stellar winds and ionizing fluxes, massive stars drive the evolution of their host environments, providing both mechanical and chemical feedback to their surroundings \citep[for a recent review, see e.g., ][]{Bresolin2008}. Binary interactions play an important role in the evolution of these massive stars: In the Milky Way, most massive stars are found in binary systems with orbital periods of less than four years \citep{Sana2011}.  These binary systems are close enough to interact during their lifetimes through mass exchange or coalescence \citep{Podsiadlowski1992}.  It is estimated that as many as 40\% of all O-type stars interact with a companion before leaving the main sequence, and more than half of these interactions (24\% of all O-type stars) lead to an overcontact configuration, where both components overfill their Roche Lobes \citep{Sana2012}.  

These interactions can fundamentally change the evolution of massive stars; however, what exactly happens during the overcontact phase is not understood well. The complex combination of physical processes occurring simultaneously during the interaction phase, including  mass exchange, internal mixing, and internal structure adjustments cause large uncertainties in evolutionary models \citep{Pols1994, Wellstein2001, deMink2007}.  Depending on the internal mixing processes, the components of a massive overcontact system can either merge or shrink within their Roche-lobes and continue to evolve via the chemically homogeneous evolution pathway \citep{Langer2012, deMink2016, Marchant2016, Mandel2016}.  Uncertainties in the evolution of these systems is further exacerbated by the lack of observational constraints: So far only eight overcontact systems are known \citep{Almeida2015, Lorenzo2014, Lorenzo2017, Hilditch2005, Howarth2015, Popper1978, Penny2008}.

In genera, despite its importance, the nature of internal mixing in massive stars is poorly understood \citep{Grin2017, Brott2011a, Bowman2019}. Constraints on these mixing processes are thus of vital importance to our understanding of massive star evolution. Determining accurate surface abundances requires high signal-to-noise spectra and appropriate nonlocal thermodynamic equilibrium (NLTE) radiative transfer codes \citep[e.g., ][]{Grin2017}. A common feature seen throughout the life cycle of a massive star is a deviation from spherical symmetry.  Even in single massive stars, deviations from spherical symmetry are common due to rotational distortion. So far, however, most of the tools currently available to spectroscopically analyze these systems are one-dimensional \citep{Hillier1998, Grafener2002, Hamann2003, Puls2005, Sander2015}. In order to properly analyze the photospheric and wind components of systems in the temperature regime where O-type stars are found, nonlocal thermodynamic equilibrium (NLTE) radiative transfer codes are required. Unfortunately, even in one dimension, these codes are quite computationally expensive, making a full three-dimensional implementation difficult when considering current computational limitations.  

One method that can help to better account for the three-dimensional nature of massive stars uses a spectroscopic surface patch model \citep{Palate2012}.  Patch models work by breaking the surface of a system into small patches and handling the spectral contribution of each patch individually.  Based on the local surface parameters, spectral profiles can be assigned and a final spectrum can be integrated by combining the individual patches across the visible surface.  This method has shown great promise in its ability to model a variety of nonspherical systems and effects \citep{Palate2012, Palate2013b, Palate2013a}.  Here we take a step further and bridge two state-of-the-art codes in order to build a three-dimensional atmospheric patch modeling tool that is both suitable for massive OB stars and broad enough to be applicable in a number of stellar and orbital configurations.

In this paper we present the \spamms (Spectroscopic PAtch Model for Massive Stars) code, a new spectroscopic analysis technique tailored to massive stars that takes three-dimensional distortion effects into account. The paper is organized as follows. Section 2 details the steps used to create the patch model and explains the input grid preparation.  In Sec. 3 we benchmark the code against \fw for spherical cases.  Section 4 provides several example applications of the code for both binary and single star cases.  In Sec. 5 we compare \spamms with similar codes and we discuss the assumptions and limitations of the code.  Finally, in Sec. 6 we summarize and discuss future plans.

\section{Method}

\begin{figure*}
   \centering
   \includegraphics[width=1\linewidth]{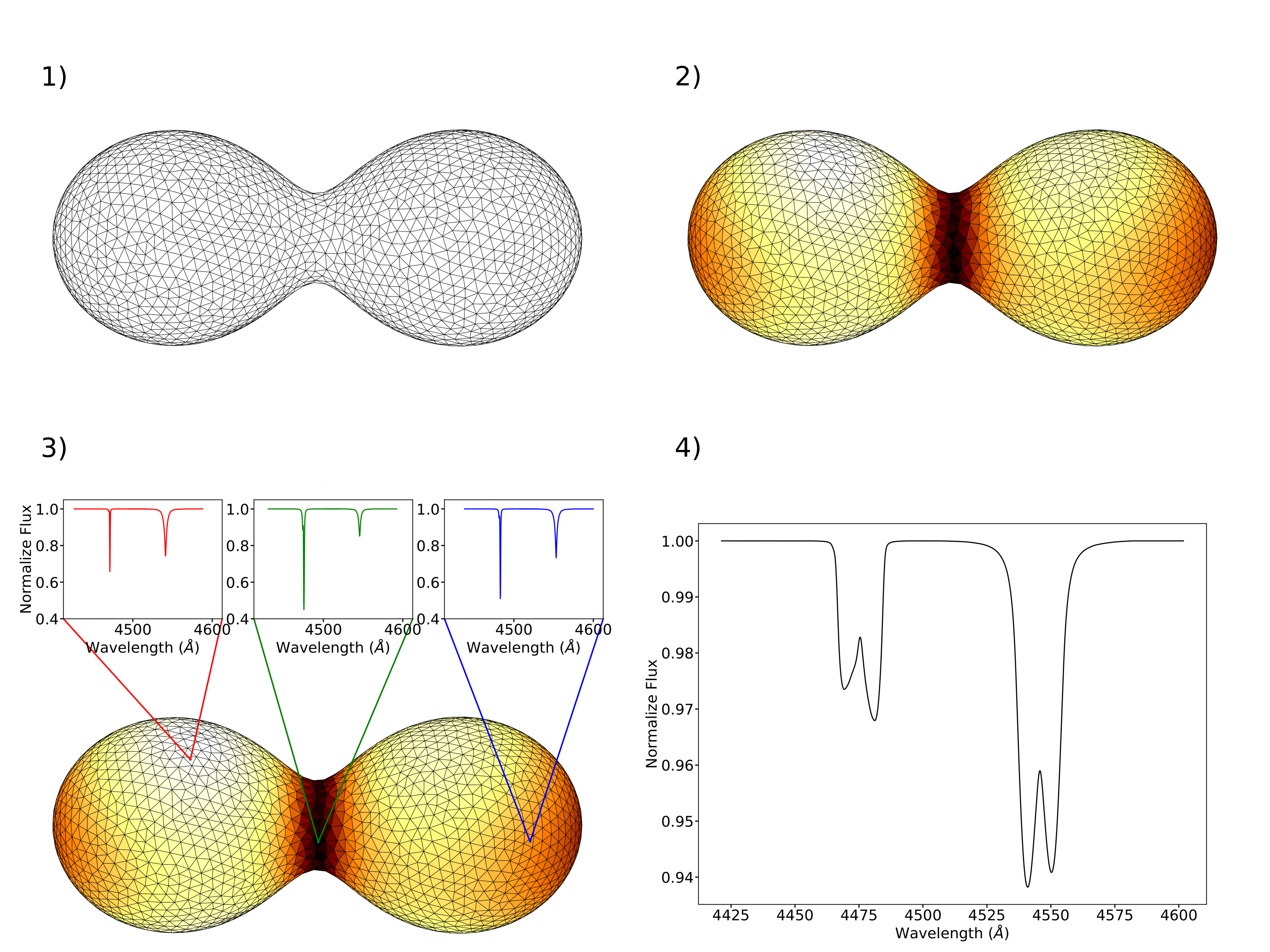}
      \caption{
      Example of the four steps of the patch model computation for an overcontact system.  Top left: A mesh representing the geometry of the system is plotted.  Top right: The same mesh is plotted, but with the face color representing the local temperature of each mesh element.  Lighter shades represent higher temperatures and darker shades represent lower temperatures.  Bottom left: Same as top right, but with assigned spectral profiles plotted for three triangles across the surface.  The \hea~\l4471 and \heb~\l4541 lines are plotted.  Bottom right: Final integrated spectral line profiles for the entire visible surface for the orientation shown in the other panels of this figure.  Again, the \hea~\l4471 and \heb~\l4541 lines are plotted.
              }
         \label{patch_steps}
   \end{figure*}

	The computation of a surface patch model can be divided into four major steps: 1) creation of a mesh representing the geometry of the system, 2) population of the mesh with local parameters, 3) assignment and Doppler shifting of spectral line profiles based on the local parameters, and 4) integration of all line profiles across the surface.  A visual representation of these steps can be found in Fig. \ref{patch_steps}.
	
\subsection{Computation and population of a geometric mesh}

	The first two steps in the computation of the surface patch model, the creation and population of a mesh, are also important for the field of eclipsing binary light curve modeling and therefore there are several codes available that focus on accurate and robust mesh creation and population \citep{Wilson1971, Wilson1979, Wilson2008, Wilson2014, Orosz2000, Prsa2005, Prsa2016}. Instead of writing our own, we choose to use \textsc{phoebe ii} (hereafter \textsc{phoebe}), one of these preexisting, well-established codes \citep{Prsa2016}. 
	
    \phoebe is a Wilson-Devinney-like code designed for the analysis and modeling of eclipsing binary systems \citep{Prsa2016}.  This code computes a forward model synthetic light curve by creating a triangulated mesh grid representing the surface geometry of each star in the system.  \phoebe allows for deviations from spherical geometry, which are taken into account via rotational distortion and/or Roche distortion. This allows for accurate modeling of nonspherical systems such as overcontact binaries, semi-detached binaries and rapidly-rotating single stars.  Given the parameters of a system and the chosen distortion method, the three-dimensional equipotential is calculated for the surfaces of the stars in the system.  The mesh is then generated by creating roughly equal area triangles where the vertices follow the equipotential surface.  These are then slightly corrected to account for the underestimation of the surface area.
    
    Once constructed, the mesh is then populated with local properties for each triangle (i.e., temperature, surface gravity, incident angle, radial velocity, etc.) taking into account a suite of physical effects including but not limited to gravity brightening, limb darkening and reflection effects.  The surface gravity per triangle is generated based on the gradient of the surface potential.  Based on the surface gravity using the \citet{vonZeipel1924} formalism, the temperature for each mesh triangle is populated. The temperature per triangle is scaled to the polar temperature which is a function of the given effective temperature.  For contact systems, the components are separated in the middle of the contact region and the generated surface elements are based solely on the component they are assigned to (e.g., when calculating the temperature, the triangle only accounts for the polar temperature of the star it is associated with).   For a detailed summary of all the physical phenomena included in \textsc{phoebe}, as well as how the mesh creation and population are computed, see \citet{Prsa2016, Horvat2018}.  The top panels of Fig. \ref{patch_steps} show these two steps of the patch model calculation for an overcontact system.

  \subsection{Emergent intensities from \fw}
	
\begin{figure*}
   \centering
   \includegraphics[width=1\linewidth]{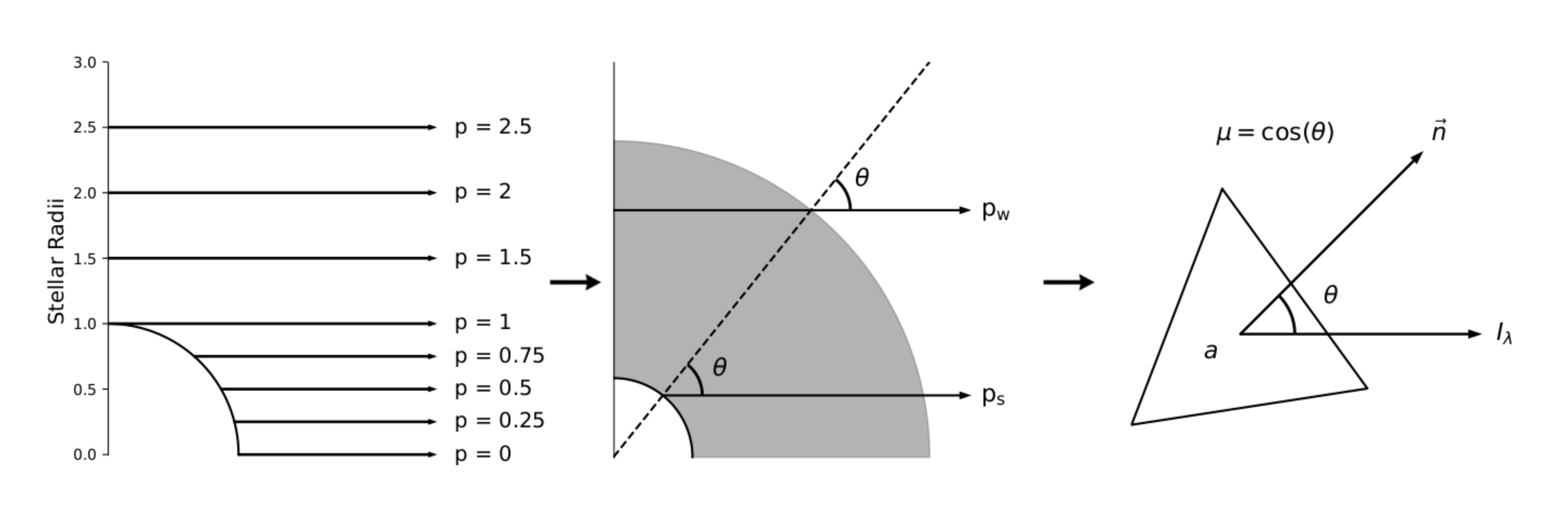}
      \caption{Left: Sketch of the p-ray geometry.  The quarter circle plotted in the bottom left represents the stellar surface and various p-rays leaving the system toward the observer (located to the right of the system) are indicated.  The p-rays with values greater than 1 originate in the wind, which extends out to 120 stellar radii.  Middle: Sketch of the conversion from p-ray geometry to emergent intensity geometry.  The quarter circle in the bottom left is the stellar surface and the shaded region represents the winds.  For a given emergent angle, two p-rays are obtained, one originating from the stellar surface ($p_s$) and one originating in the wind ($p_w$) Right: Sketch of emergent intensity geometry.  A patch of the stellar surface is plotted with two emerging rays, one normal to the surface (labeled $\vec{n}$) and one pointing to the observer (labeled $I_\lambda$).   $\theta$ is the emergent angle of the ray toward the observer and $a$ represents the area of the patch.
      }
         \label{p_ray_emergent_geo}
   \end{figure*}

	In order to properly calculate the total integrated line profiles across a stellar surface, we need to obtain the emergent intensities as a function of wavelength for each discretized surface element. The flux at each wavelength bin is given by

	\begin{equation}
      F_\lambda = \sum\limits_{n} I_{\lambda, n} a_n \mu_n v_n,
      \label{flam}
	\end{equation}
   where the subscript $n$ denotes a specific triangle in the mesh, $I_{\lambda, n}$ is the emergent intensity at a given wavelength for triangle $n$, $a_n$ is the area of the triangle, $\mu_n$ is the cosine of the emergent angle, and $v_n$ is the fraction of the triangle that is visible to the observer (this is only relevant if a portion of the triangle is obscured or if the triangle is pointed away from the observer).  The values in Eq. \ref{flam} are visually represented and labeled in the right panel of Fig. \ref{p_ray_emergent_geo}.
   	
   The emergent intensity line profiles need to be calculated with an appropriate radiative transfer code, here we adopt \fw \citep{Puls2005}.  \fw is a unified, NLTE atmosphere and spectrum synthesis code suited for stars of spectral type O, B, and A, and a wide range of wind-strengths.  It is therefore appropriate for the analysis of stars across the OB-type spectral range.  Given a set of input stellar and wind parameters, \fw first calculates the atmosphere and wind structure of the star, and then computes the formal solution, returning a user defined set of intrinsic spectral line profiles.  However the default line profiles given by \fw are integrated line profiles in units of flux, and not emergent intensities.  Thus, as a first task, we need to retrieve the emergent intensity line profiles from \fw.
   
   \fw operates using p-z geometry, which uses a series of parallel rays (denoted p-rays) that probe different regions of the stellar surface and wind \citep{Hummer1971}(see left panel of Fig. \ref{p_ray_emergent_geo}).  These p-rays are denoted by their distance from the center of the star on their closest approach in units of stellar radii (i.e., the p-coordinate defines the so-called impact parameter).  A ray that passes through the center of the star is given a p-ray value of 0 while one that is tangential to the surface of the star has a p-ray of 1 and a ray passing through the wind has a p-ray of greater than 1.  Using the stellar and wind structure, \fw creates 64 p-rays by default and performs radiative transfer calculations for each of them.  As mentioned above, these are then integrated to get the final flux profile. Here we instead use these p-rays to obtain the emergent intensities.  Since the p-rays probe different regions, each p-ray corresponds to a specific emergent intensity at a specific emergent angle and thus a specific value of $\mu$.  It is a trivial exercise to interpolate across each wavelength point in the p-rays to obtain the emergent intensity as a function of $\mu$ and wavelength for both the photosphere and the wind.  Figure \ref{p_ray_emergent_geo} shows a comparison of the p-ray geometry and the emergent intensity geometry.
   
   The mesh created by \phoebe only corresponds to the photosphere of the system, while the winds are extended and optically thin.  Optically thin winds do not eclipse in the same way that a photosphere does, so to account for this difference, we compute two separate sets of emergent intensities, one that corresponds to the photosphere of the star and one that corresponds to the extended winds. To limit the complexity, we use the same mesh geometry for the photosphere and the winds (the effects of this assumption are discussed in Sect. \ref{assumptions}).  For our computations, the wind extends out to $\sim$ 120 stellar radii.  Therefore, for the calculation of the emergent intensities, p-rays originating from the stellar surface (i.e., 0 -- 1) are used for the photosphere and to avoid double counting, p-rays originating in the wind (i.e., 1 -- 120)  are used for the winds.  Since we do not include p-rays down to 0 for the winds, the maximum $\mu$ does not reach 1, so we rescale the $\mu$ angles such that a p-ray of 1 corresponds to a $\mu$ of 1.  We then interpolate and create a set of 101 emergent intensity line profiles for the photosphere and wind corresponding to $\mu$ angles between 0 and 1 inclusive in steps of 0.01.
   
  \subsection{Computation of input \fw grid}
   
\begin{figure}
   \centering
   \includegraphics[width=1\linewidth]{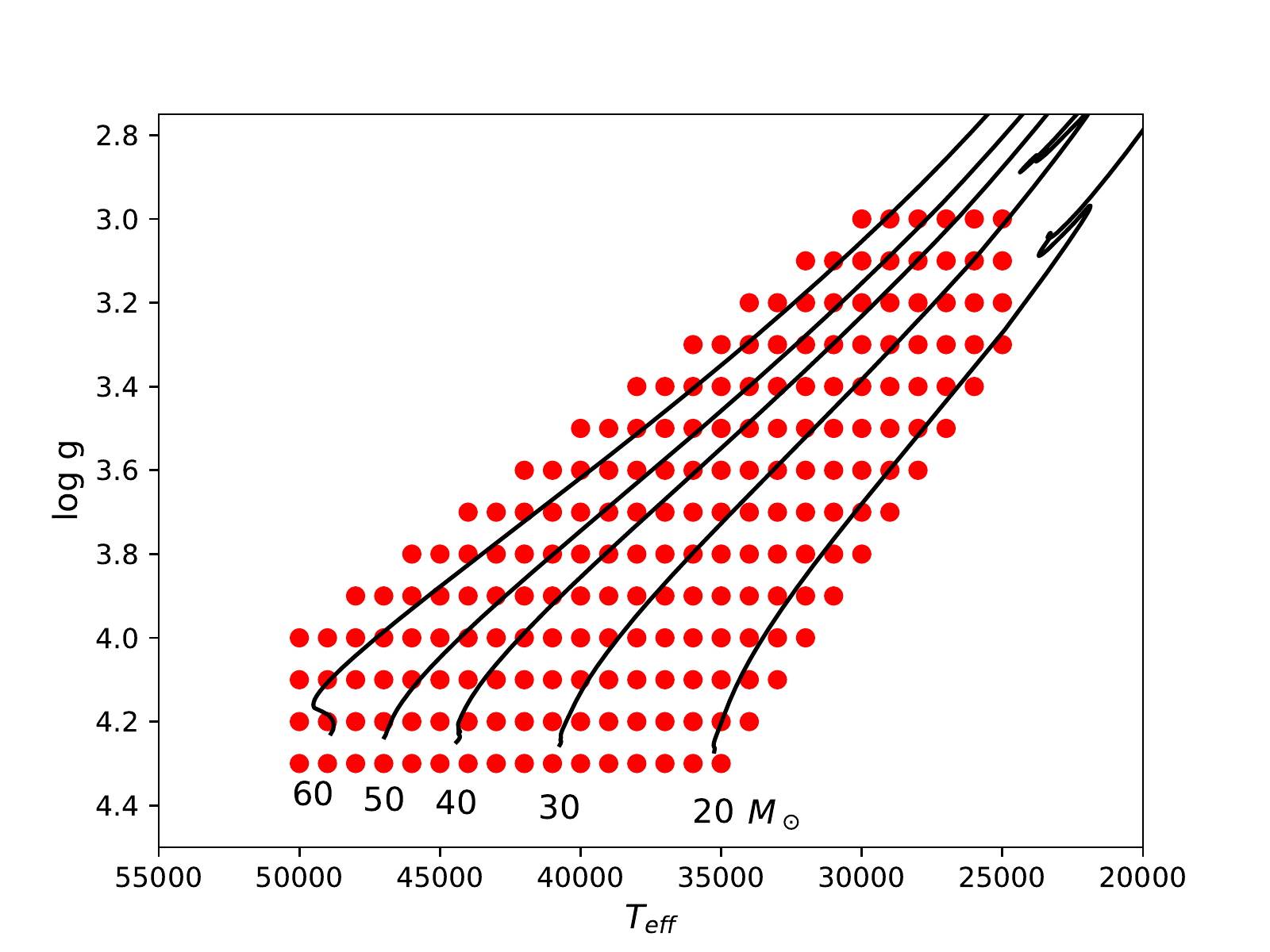}
      \caption{Coverage of \fw grid in temperature surface gravity space.  Evolutionary tracks from \citet{Brott2011a} are plotted for various masses in black and our grid points are represented by the red dots.
              }
         \label{spec_hrd}
   \end{figure}

   We can now use this method to create a grid of \fw models, which is then used for the assignment of spectral line profiles to the mesh. The grid of \fw models covers a 5 dimensional parameter space: effective temperature, surface gravity, radius, Helium abundance and CNO abundance.  The temperature and surface gravity dimensions cover most of the main sequence lifetime of typical massive stars between the masses of 20 and 60 \msun.  Figure \ref{spec_hrd} shows the coverage of the \fw grid for the temperature and surface gravity dimensions, which are in steps of 1,000 K and 0.1 dex respectively.  The grid also covers 11 radius points, spanning from 6.5 to 9.0 \rsun\  inclusive in steps of 0.25 \rsun\  and 4 helium abundance steps: 0.06, 0.10, 0.15 and 0.20, where the helium abundance ($Y_\mathrm{He}$) is given by:
   \begin{equation}
      Y_\mathrm{He} = \frac{N_\mathrm{He}}{N_\mathrm{H}}.
   	\end{equation}	
	We note that $N_\textrm{He}$ and $N_\textrm{H}$ are the number abundances of helium and hydrogen respectively.  Finally, the grid includes 5 CNO abundance steps: 6.5, 7.0, 7.5, 8.0 and 8.5, where the abundance per element is given by: 
	\begin{equation}
      \varepsilon_\mathrm{X} = \log \frac{N_\mathrm{X}}{N_\mathrm{H}} + 12.
   	\end{equation}
   	Here, $N_\textrm{X}$ and $N_\textrm{H}$ are the number abundances of the given element and hydrogen respectively.  The grid is calculated with $\varepsilon_\mathrm{C} = \varepsilon_\mathrm{N} = \varepsilon_\mathrm{O}$, however while equal abundances are indeed unphysical, this does not have a large effect on the CNO lines themselves, the lines of other elements or the atmospheric structure \citep{Carneiro2019}. The grid is computed at LMC metallicity ($Z_\mathrm{LMC} =0.5\ Z_\odot$) with wind parameters typical for this mass range and metallicity.  We assume a terminal wind speed of 3000 \kms and a velocity field exponent $\beta$ of 0.8.  The mass loss rate is calculated using the prescription described in \citet{Vink2001} with an assumed mass of 30\msun{}. This mass choice is motivated by the derived component masses of an overcontact binary discussed in Sect. \ref{352}, however the code allows for the usage of a user-defined input grid. In the present case, the resulting grid contains 43,780 \fw models.  This is sufficient for our purposes but this grid can easily be extended if required in the future.

\subsection{Assignment and Integration of line profiles}
The third step in the calculation of the surface patch model is the assignment of the spectral line profiles to the mesh.  As mentioned above, the mesh created and populated by \phoebe contains the local properties needed to assign the \fw models, namely the temperature, surface gravity, radius and incident angle.  The temperature and the incident angle affect the final line profile much more significantly than the radius or surface gravity do as they both directly influence the emergent intensity.  For this reason, we choose to interpolate between grid points for these two parameters while the surface gravity and radius for each mesh point are rounded to the nearest value contained in the grid.  Thus, the final emergent intensity line profile for each mesh point is generated by interpolating between the four emergent intensity line profiles corresponding to the closest two temperatures and incident angles in the grid. Each profile is then Doppler shifted based on the local radial velocity of that mesh point.  This is done for both the photosphere and the wind component so each mesh point contains two separate line profiles.

The final step is the integration of the line profiles.  Each mesh point now contains two wavelength and two intensity arrays corresponding to its local parameters, however the wavelength arrays between different mesh points are different from one another due to the Doppler shifting.  We thus resample all line profiles on the mesh to the same wavelength array using linear interpolation.  The final integration for the photospheric component is calculated using Eq. \ref{flam}, where the area ($a_n$), cosine of the emergent angle ($\mu_n$) and visibility ($v_n$) are given in the mesh.  The wind component is calculated in the same way with the caveat that the visibility is not used.  This is because the visibility defines which mesh points are being eclipsed, but the wind is optically thin and thus it does not eclipse itself.  Additionally, a scaling factor is added to account for the flux difference that occurs when collapsing the winds to the surface of the star.  The integrated photosphere and wind profiles are added together and normalized so the final output is a normalized flux line profile.  Finally, if any wind mesh component contributes more than 5\% of the total flux of the system, it is removed and the final flux line profile is recalculated.

\section{Benchmarks}
Before applying the patch model to systems with nonspherical geometry, we first benchmark its performance against \fw for three spherical test cases. These test cases are detailed below.

\begin{figure*}
   \centering
   \includegraphics[width=1\linewidth]{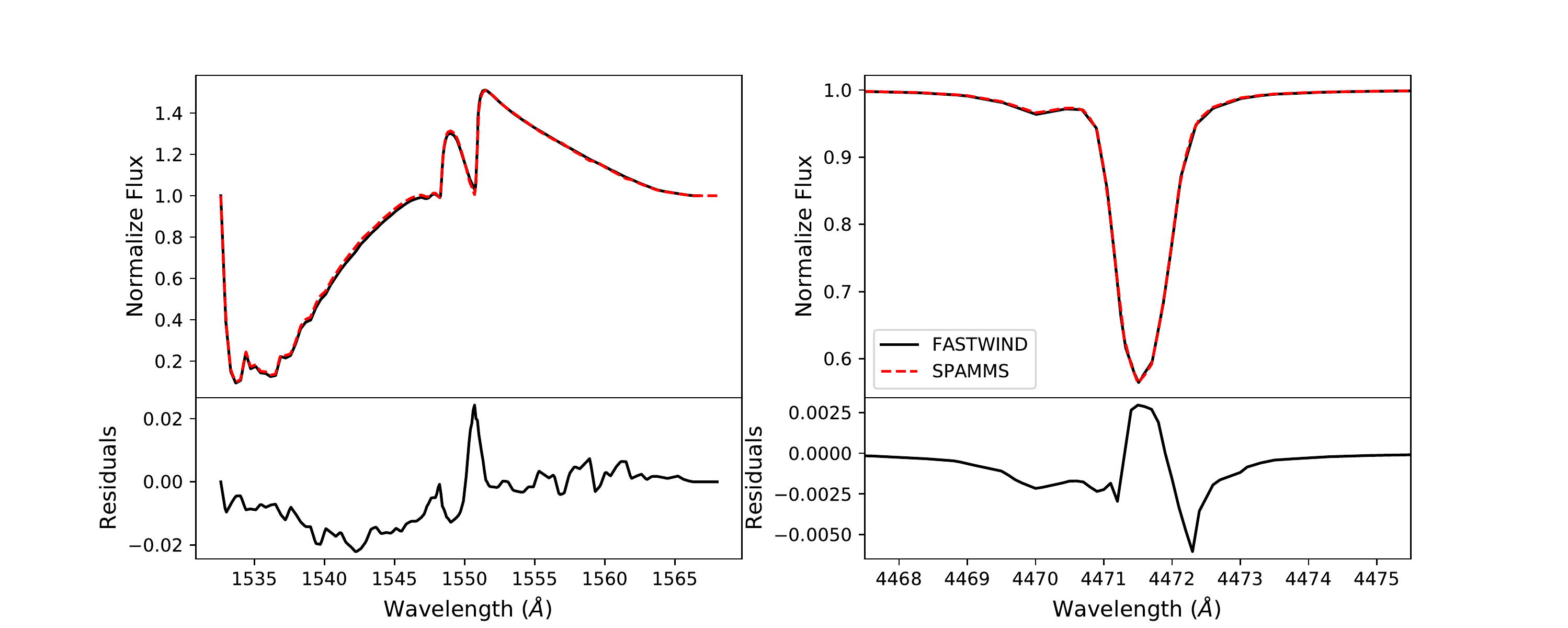}
      \caption{Comparison between the integration method used by SPAMMS and \fw for a system with the same parameters. SPAMMS line profiles are plotted in red dashed lines and \fw line profiles are plotted in black, for a wind line \cd\ \l1548 on the left and a photospheric line \ha\ \l4471 on the right.  The residuals between the two methods are plotted beneath each.
              }
         \label{benchmark1}
   \end{figure*}

The goal of the first test case is to ensure that the conversion from p-ray geometry to emergent intensity geometry and the subsequent integration is consistent with \fw.  Since our goal is to test the integration, we chose a point which falls exactly on one of our grid points to avoid any uncertainties that may arise from interpolating between grid points. We compute a nonrotating, spherically-symmetric SPAMMS model of a 42,000K star with a radius of 7.5\rsol, a surface gravity of 4.2 (which corresponds to a mass of $\sim$ 32\msol), a helium abundance of 0.1 and a CNO abundance of 7.5.  A \fw model with the same parameters is computed and we compare the spectral line profiles of two lines: \cd~\l1548,1550 a doublet wind line that has a relatively strong P-Cygni profile and \hea~\l4471 a photospheric line that is not strongly affected by the winds.  For both lines, we consider only the mentioned transitions and do not include other transitions that may fall in the spectral region.  Figure \ref{benchmark1} shows this comparison and the corresponding residuals between the SPAMMS model and the \fw model.  For the wind line, the residuals show an agreement within 1\% for a majority of the profile with a spike up to 3\% for the complex core of the line.  The photospheric line shows an agreement within a third of a percent in the core and less than a tenth of a percent in the wings of the line.  Overall, these results show that the integration method for SPAMMS is in very good agreement with \fw.

\begin{figure*}
   \centering
   \includegraphics[width=1\linewidth]{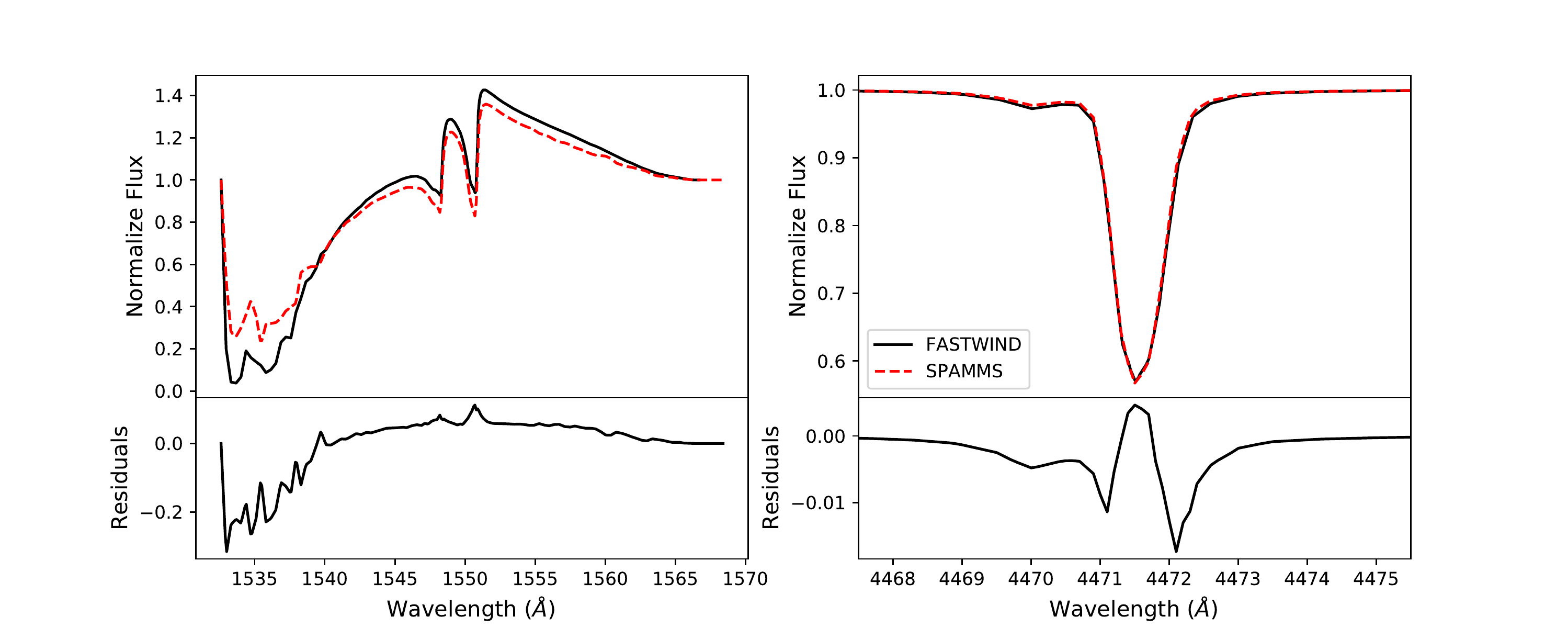}
      \caption{Comparison between line profiles generated using grid interpolation with SPAMMS and  line profiles generated with \fw for a system with the same parameters. SPAMMS line profiles are plotted in red dashed lines and \fw line profiles are plotted in black, for a wind line \cd\ \l1548 on the left and a photospheric line \ha\ \l4471 on the right.  The residuals between the two methods are plotted beneath each.
              }
         \label{benchmark2}
   \end{figure*}

The goal of the second test case is to ensure that the linear interpolation between grid points is reasonable when compared to the exact solution given by \fw for the same model.  To do this, we chose a parameters that are in the exact middle between grid points since this represents the farthest deviation possible.  We compute a nonrotating spherically-symmetric SPAMMS model of a 42500K star with a radius of 7.125\rsol, a surface gravity of 4.15 (corresponding to a mass of $\sim$ 26\msol), a helium abundance of 0.1 and a CNO abundance of 7.5.  We do not choose a point between abundance bins because the step sizes in abundance have much larger effects on the final line profiles than the other parameters.  While the code only interpolates between temperature grid points, we deviate in surface gravity and radius as well to show a worse-case scenario.  Figure \ref{benchmark2} shows the comparison between the SPAMMS and \fw line profiles and the residuals between the two models.  For the wind line, the shape of the line is reasonable and for most of the profile, the two models agree within a few percent, however the maximum deviation reaches almost 30\% at some points.  The photospheric line on the other hand shows an agreement within 1\% for most of the line with a maximum deviation of less than 2\%.  These results show that the linear interpolation works very well for photospheric lines, but does not have the same accuracy for wind lines.

\begin{figure*}
   \centering
   \includegraphics[width=1\linewidth]{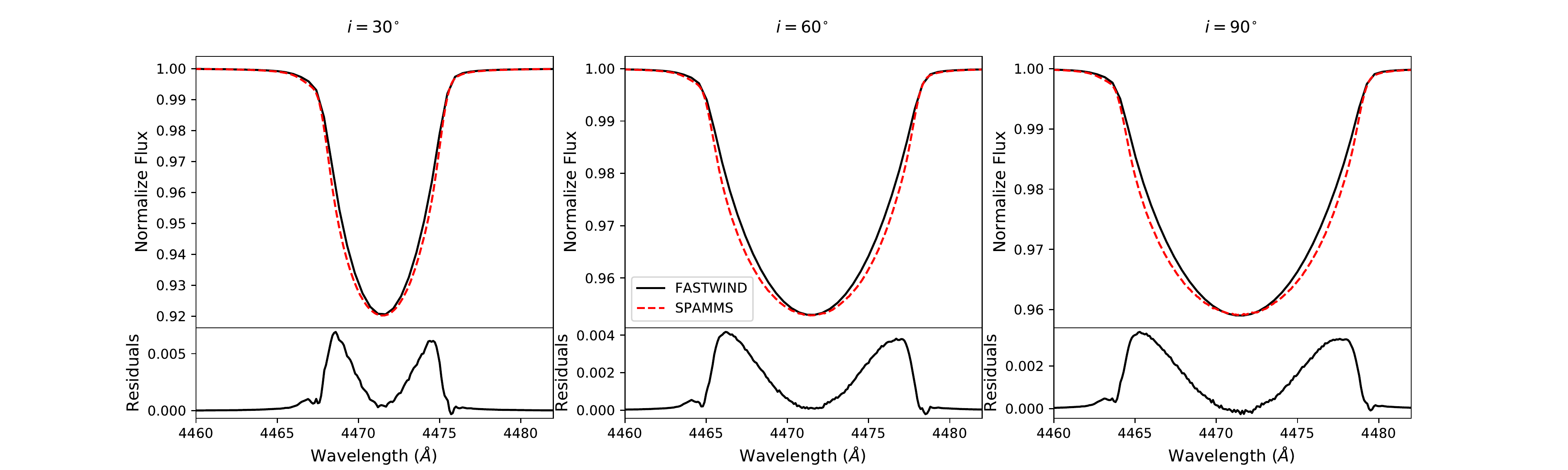}
      \caption{Comparison between line profiles from a rotating star generated using SPAMMS and  \fw for a system with the same parameters. SPAMMS line profiles are plotted in red dashed lines and \fw line profiles are plotted in black, for the photospheric line \ha\ \l4471.  The residuals between the two methods are plotted beneath each.
              }
         \label{benchmark3}
   \end{figure*}

The goal of the third test case is to ensure that the effects of rotation are properly accounted for in the patch model during the computation of the final line profiles. To do this, we compute a rotating \spamms model with spherical geometry (rotational distortion is switched off) and a corresponding \fw model with the same parameters, which is rotationally broadened via convolution with a rotation profile to the corresponding projected rotational velocity.  We use the same stellar parameters as for the first benchmark case and compute three models with the same rotation rate (70\% critical), but different inclinations ($30^{\circ}$, $60^{\circ}$ and $90^{\circ}$).  Figure \ref{benchmark3} shows the comparison between the two methods.  The line profile computed by \spamms is slightly different from that calculated by \fw and broadened, but the two are consistent within half of a percent.

\section{Applications}
While our initial goal was to model overcontact systems, the patch model that we developed has a broader range of applicability.  Anything that \phoebe can model and that falls within the range of parameters suitable for \textsc{fastwind}, can be analyzed with the patch model. A few examples are discussed below.

\subsection{Overcontact binaries} \label{352}
Overcontact binaries are arguably the most extreme example of nonspherical systems that would benefit from the patch model.  With their unique shapes, the difference in temperature across the surface of massive overcontact systems can be on the order of 10,000K or more.  Despite this, the way that these systems are spectroscopically analyzed currently involves spectral disentangling followed by atmosphere fitting of each component individually, while the effect of the bridge is omitted completely.  As discussed in \citet{Abdul-Masih2019}, this analysis method assumes spherical symmetry several times.  The disentangling process not only assumes spherical geometry, but it also blends any profile variations across the surface, returning an averaged spectrum for each component in the system. Furthermore, the determination of the atmospheric parameters also assumes spherical symmetry as the available atmosphere codes are all one-dimensional.

With the patch model, we can better represent the three-dimensional surface of these systems.  As an example, we analyze the massive overcontact binary VFTS 352, which has been previously analyzed both photometrically and spectroscopically \citep{Almeida2015, Abdul-Masih2019}.  VFTS 352 is an eclipsing double-lined spectroscopic binary with masses of $\sim$ 29 + 29\msol, radii of 7.22 + 7.25\rsol, temperatures of 42,540 + 41,120K and an orbital period of 1.124 days \citep{Almeida2015}. The system is in deep contact with a fillout factor of approximately 1.29, causing large variations in surface gravity and temperature across the surface: $\sim$ 3.7 to 4.3 and $\sim$32,300 to 45,800K respectively when assuming the solution obtained from \citet{Almeida2015}.  \citet{Walborn2014} derived spectral types of O4.5 V(n)((fc))z +O5.5 V(n)((fc))z) for this system.  In both \citet{Almeida2015} and \citet{Abdul-Masih2019} it was shown that the component stars were hotter than expected for stars of their mass and rotation rate indicating an extra source of mixing.  The spectroscopic analysis was done using spherical models, so we can investigate whether a 3-dimensional analysis using the patch model yields a similar result.

\begin{figure}
   \centering
   \includegraphics[width=1\linewidth]{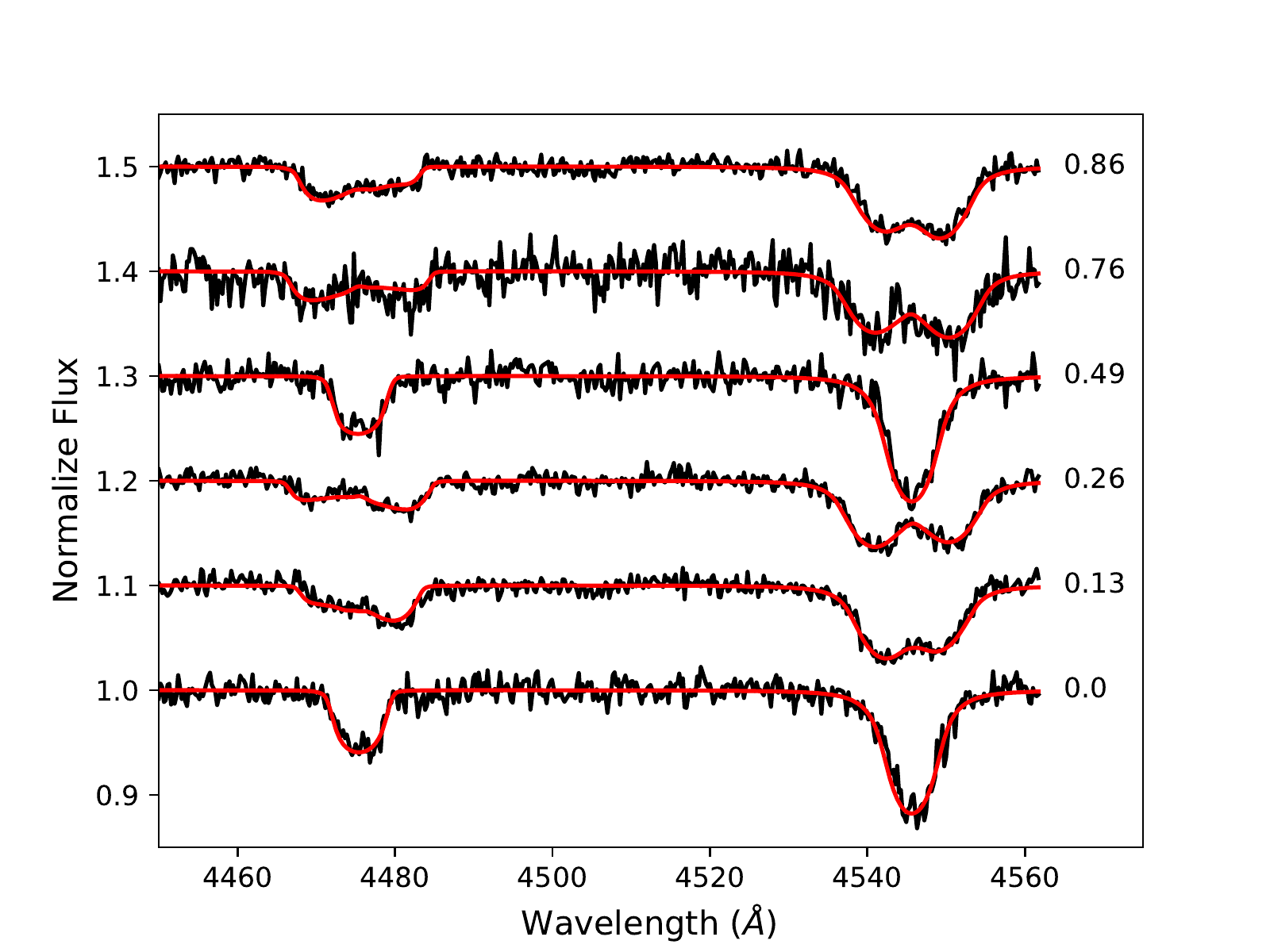}
      \caption{Best fit model for VFTS 352 shown at various phases.  The observed spectra are plotted in black while the model is overplotted in red.  The phase is given to the right of each model.
              }
         \label{vfts_352}
   \end{figure}

Using a grid-search chi-square minimization routine, we attempt to determine the effective temperature of both the primary and secondary using the patch model.  We use the photometric solution (period, inclination, mass ratio, fill-out factor, and semi-major axis) from \citet{Almeida2015} and the 32 epochs of the optical FLAMES spectra described in \citet{Abdul-Masih2019} as inputs. Assuming a temperature distribution across the surface as described by \citet{vonZeipel1924}, we then fit the line profiles of \hea~\l4471 and \heb~\l4541 (two lines typically used to constrain the temperature) to each phase of the observed data. The system is known to have weak winds, and since the input grid assumes a \citet{Vink2001} mass loss prescription, we use lines that are not affected strongly by the winds for the fitting.  We fit all 32 phases simultaneously to obtain a global best fit for the temperatures of the two components. The resulting best-fit temperatures for the primary and secondary were 44,000 and 41,400K respectively, which agree within error to the values determined in \citet{Abdul-Masih2019}.   Figure \ref{vfts_352} shows the comparison between the observed data and the line profiles generated using the best-fit for several different phases.

\subsection{Rapidly rotating single stars}
As stated above, currently all NLTE radiative transfer codes for massive stars are one dimensional and thus assume spherical geometry.  Rotational distortion, however can have significant effects on the temperature profile across the surface of a massive star \citep{vonZeipel1924}.  As a star rotates more rapidly, it begins to oblate; and due to the change in geometry, the poles experience a higher effective surface gravity and higher temperature while the equator experiences a lower effective surface gravity and lower temperature.  Thus, if a rapidly rotating star is inclined toward the observer, the star appears hotter than if viewed with an inclination of 90.  This effect is not taken into account in spherically symmetric codes, however with the patch model, we can investigate this effect.

\begin{figure*}
   \centering
   \includegraphics[width=1\linewidth]{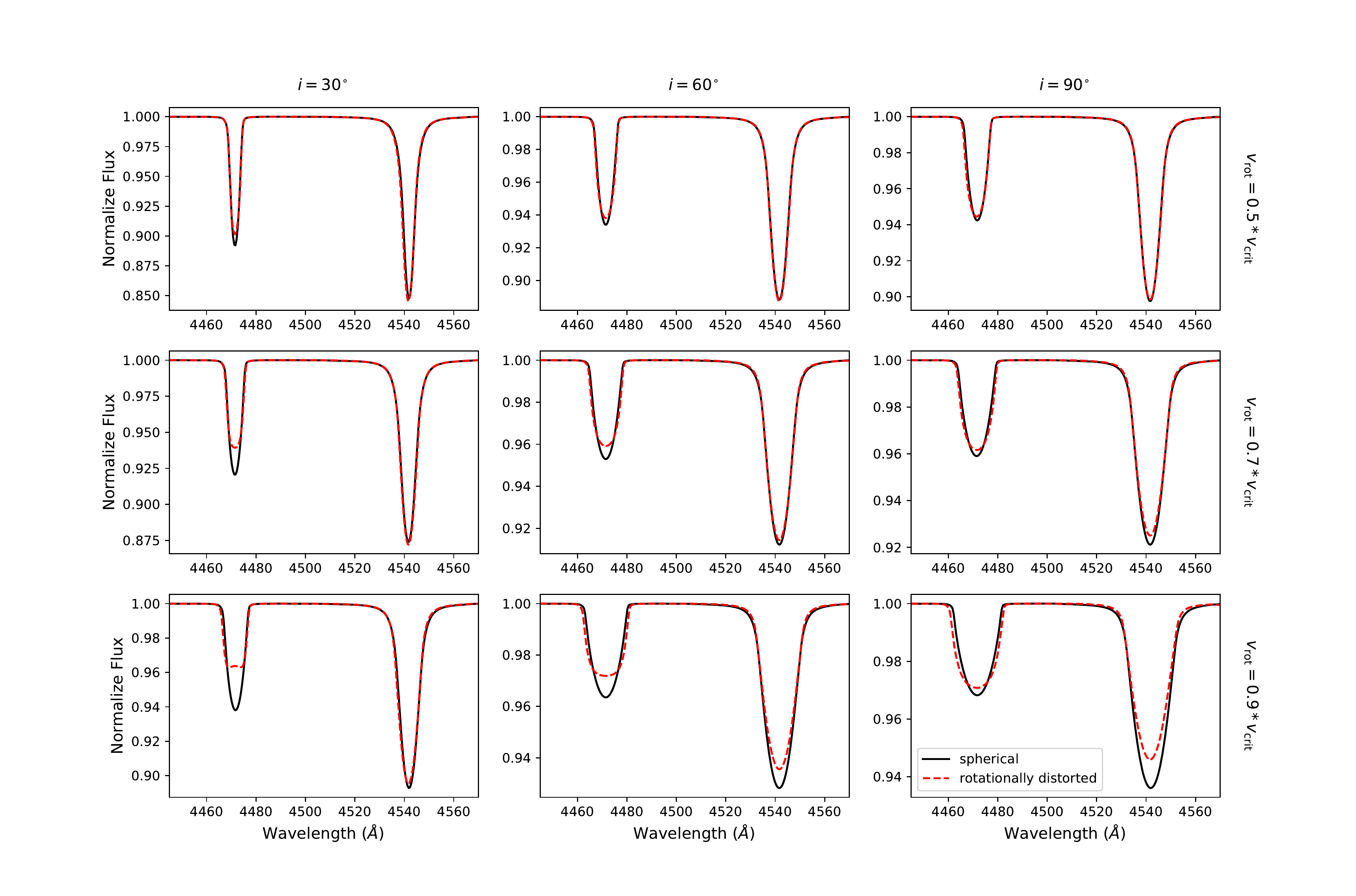}
      \caption{Comparison of \hea~\l4471 and \heb~\l4541 line profiles when assuming spherical geometry and rotational distortion for a rapidly rotating system in black and red respectively.  Three different rotation rates (indicated on the right of the figure and three different inclinations (indicated above the figure) are plotted for comparison.
              }
         \label{RR_temp}
   \end{figure*}

Consider a 42,000K, 30\msun\ star with an effective radius of 7.5\rsun.  Using SPAMMS, we model the system at three different rotation rates (50\%, 70\% and 90\% critical) and inclinations ($30^{\circ}$, $60^{\circ}$ and $90^{\circ}$) to see how the rotational distortion impact the line profiles. We calculate two line profiles for each rotation-rate-inclination combination: \hea~\l4471 and \heb~\l4541.  Figure \ref{RR_temp} shows line profiles for these combinations assuming both rotationally-distorted geometry and spherical geometry.  For the 50\% critical case, the line profiles appear almost identical at an inclination of 90 degrees.  As the inclination decreases however, the rotationally distorted model begins to deviate from the spherical model.  The \hea~\l4471 line begins to appear weaker while the \heb~\l4541 line appears slightly stronger.  Observationally, this would indicate a higher temperature.  This trend becomes more apparent when considering the 70\% critical case, where a similar effect can be seen.  In the 90\% critical case, the inclination effect is more apparent still.  Other effects begin to present themselves in this case however.  At an inclination of 90 degrees, the rotationally deformed model appears cooler than the spherical case, while the two models only appear to show the same temperature at an inclination of 60 degrees.  Interestingly, both helium lines are weaker at 60 degrees, observationally indicating a lower helium abundance.  Finally, at an inclination of 30 degrees, the \hea~\l4471 line begins to show a double peak.  Since the hotter pole has a relative radial velocity close to 0, it is contributing significantly to the center of the \hea~\l4471.  The increased temperature however means that the \hea\ lines are weaker than the cooler portions of the surface, thus causing this double-peaked feature.

With the exception of the 90\% critical 90 degree inclination model, all of the others show a slightly larger \heb\ to \hea\ ratio than their spherical counterparts indicating that the system appears hotter than the spherically symmetric model.  This implies that spectroscopically-determined temperatures using spherically symmetric models are systematically overestimated for rotating systems.  The full extent of these nonspherical effects and their implications will be investigated in a future paper.


\subsection{Rossiter-McLaughlin Effect}
\begin{figure*}
   \centering
   \includegraphics[width=1\linewidth]{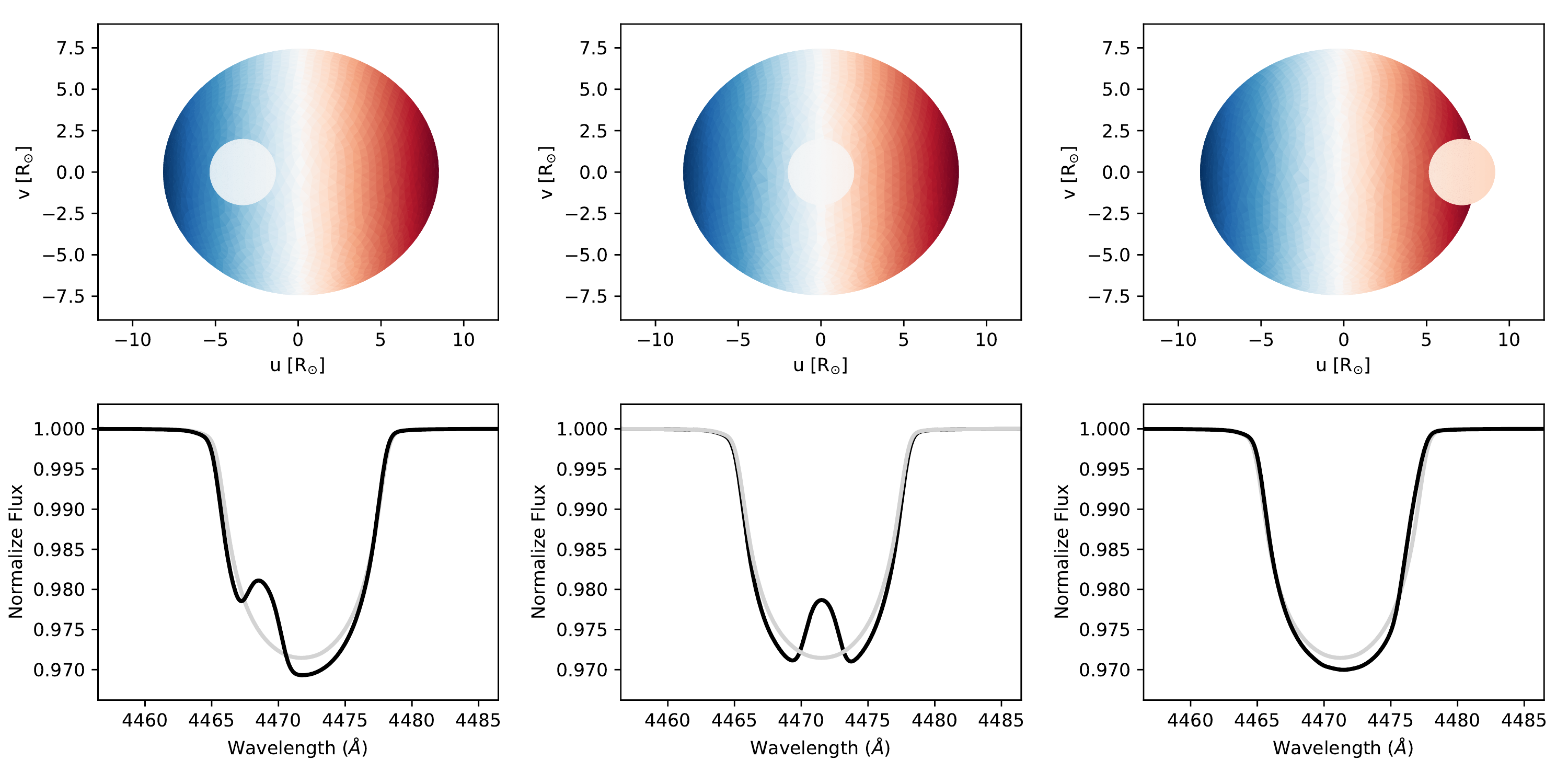}
      \caption{Example of the Rossiter-McLaughlin effect.  Top panels show the rapidly rotating star at different points during the eclipse.  The face-color represents the radial velocity, with blue regions being blue-shifted and red regions being red-shifted.  The bottom panels show the modeled \hea~\l4471 spectral line plotted in black at the same phase as the respective top panel and the gray line represents the out of eclipse line profile.
              }
         \label{rossiter}
   \end{figure*}

Another effect that can be studied with the patch model is the Rossiter-McLaughlin effect. As a star rotates, regions of the star are blue shifted while other regions are red shifted.  If this star is eclipsed then its observed rotational profile changes since portions of the blue or red shifted regions are no longer visible.  This change in the rotational profile in turn changes the appearance of the observed spectral lines. To demonstrate this effect, we model a theoretical system with component radii of 8\rsun\ and 4\rsun, temperatures of 45,000K and 20,000K, a period of 5 days, a semi-major axis of 40\rsun\ and a mass ratio of 0.2.  The system has an inclination of 90$^\circ$ and the primary is rotating asynchronously with a period of 1 day. Figure \ref{rossiter} shows how the spectral line profiles change as the primary is being eclipsed. These line profile variations can be used to determine if the axis of rotation is misaligned from the orbital axis.

\subsection{Struve-Sahade Effect}
\begin{figure}
   \centering
   \includegraphics[width=1\linewidth]{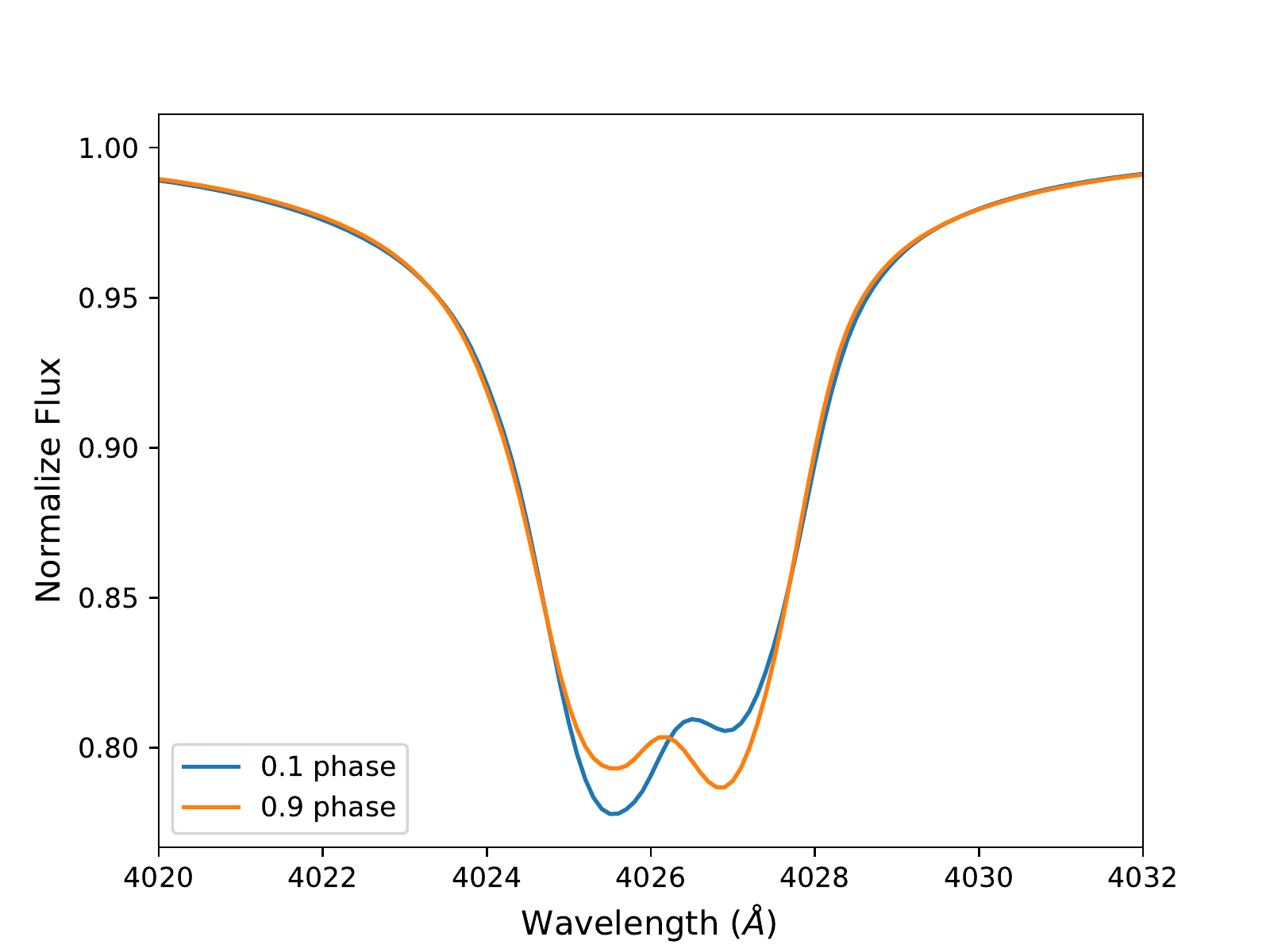}
      \caption{Example of the Struve-Sahade effect for a noneclipsing spectroscopic binary.  The \hea~\l4026 line profile is plotted for the system at two different phases (0.1 plotted in blue and 0.9 plotted in orange) demonstrating the change in line strengths.
              }
         \label{struve-sahade}
   \end{figure}

The Struve-Sahade effect is an effect observed in spectroscopic binary systems where some of the spectral lines of the secondary appear to strengthen while blue-shifted and weaken while red-shifted \citep{Struve1937}.  Several possible explanations for this effect have been proposed (i.e., gaseous streams obscuring the secondary, wind-wind collisions deflected by the Coriolis force, surface flows, etc.) however it is still unclear which mechanism is the cause \citep{Struve1950, Sahade1959, Gies1997, Gayley2007, Linder2007}.  \citet{Palate2012} demonstrated that this effect could be replicated using a patch model approach similar to \spamms.  To see if we too can reproduce the Struve-Sahade effect, we compute a \spamms model based on the massive binary HD 165052, a noneclipsing system that is known to display this effect \citep{Linder2007}. This system has component equivalent radii of 9.29\rsun\ and 8.65\rsun\ respectively, temperatures of $\sim$ 35,000K and 34,000K, a period of 2.95515 days, a semi-major axis of 31.25\rsun\ an inclination of 23$^\circ$ and a mass ratio of 0.87 \citep{Linder2007}. The radii of this system place it outside of our computed grid but with slight adjustments to the parameters, we can compute a model for a similar, theoretical system. We adjust the radii and semi-major axis by setting them equal to 90\% of their calculated values and adjust the period to ensure that the masses remain the same. Assuming synchronous rotation, we model the \hea~\l4026 line at two different phases (0.1 and 0.9) and find that we are indeed able to reproduce the Struve-Sahade effect for this system.  Figure \ref{struve-sahade} shows a comparison of these two phases.  While our parameters are slightly different from those cited in \citet{Palate2012}, the line profiles appear in very good agreement with each other.  The fact that the patch model can reproduce the Struve-Sahade effect implies that it may originate from either a pure geometric effect or a surface temperature distribution effect.

\section{Discussion}
\subsection{Comparison to CoMBiSpeC}
While the concept of a patch model is not new, our implementation has several advantages over previous codes, such as the CoMBiSpeC code presented by \citet{Palate2012}.  There are four major features of our code that set it apart: the use of \phoebe as the base for the computation and population of the mesh, the inclusion of winds in the model, the use of emergent intensities instead of flux for the assignment of line profiles and the dimensionality of the grid from which the line profiles are assigned.

There are several advantages of using \textsc{phoebe}, one of which is the amount of physics already implemented in the code.  \phoebe uses an advanced mesh triangulation technique that not only allows it to recreate the complex geometries, but also prevents holes and overlaps which other forms of meshing sometimes suffers from.  In addition, it is constantly being updated with new physics, which we can use.  For example, \phoebe is built to eventually support triple and higher order multiple systems, which we would be able to use for the patch model \citep{Prsa2016}. Using a well established (and well tested) code like \textsc{phoebe}, we can avoid development and trouble shooting of these improvements ourselves and reach the wider community of \phoebe users.

The second major feature of SPAMMS that differs from other codes is the inclusion of winds.  For massive stars, the winds can have a major impact on the spectral line profiles.  By including winds, we can fit many important diagnostic lines both in the UV and optical portions of the spectrum that would otherwise be impossible to reproduce. 

The third major difference between SPAMMS and other codes is the use of emergent intensities when assigning line profiles to the mesh.  CoMBiSpeC uses flux line profiles that are scaled by the projected area and applies a linear limb-darkening law. By using emergent intensities however, we can better account for the change in intensity as a function of emergent angle.  Additionally, we can obtain flux-calibrated line profiles, which can be directly compared to observed data.

Another important difference is the dimensionality of our \fw grid in SPAMMS.  We include the radius of each mesh point as a constraining parameter when assigning line profiles, which is not done in \citet{Palate2012}.  We also compute a variety of helium and CNO abundance combinations allowing us to obtain abundance information.

\subsection{Assumptions and Limitations} \label{assumptions}

While this patch model is a welcome addition to the tools available for the analysis of massive stars, it is not perfect and some of the assumptions can lead to errors.  The first major source of error is the use of one-dimensional single star atmosphere models for the construction of the grid.  We assume that the local conditions at each point across the surface is equivalent to a single spherical star with the same conditions, which is not necessarily the case.  

The treatment of the wind is also not optimal.  The wind is assumed to start at a p-ray of 1, which can cause issues at $\mu$ angles very close to 1.  There is a steep drop-off in intensity in the transition region between the star and the winds.  The wind component of triangles that are viewed very close to face on may fall in this drop-off region and contribute more flux than they should.  We take this into account by removing the wind component of any triangle that is contributing more than 5 percent of the total flux but this is an arbitrary cut.  Furthermore, the collapsing of the wind to the surface of the star is not physical.  The winds are extended and are most likely much more uniform than the wind structure assumed by the patch model (i.e., the wind component of each triangle is based on the surface conditions and not on the bulk conditions of the system that is driving the wind).  

In addition to the nonuniformity of the wind, further nonphysical effects appear when considering the case of binaries and their interaction with three-dimensional wind structures.  For binary systems with strong winds, wind-wind collisions become important, however since the winds are collapsed to the surface of the star in the patch model, this effect cannot be reproduced. Furthermore, since the winds are not allowed to eclipse each other or be eclipsed by the star in our formalism, there is no way to reproduce a Rossiter-McLaughlin-like effects for the wind emission.  It is also important to note, that a correct UV-line synthesis requires that the wind-structure above the patches is represented in a realistic way; for example, the \citet{Vink2001}-scaling is not valid for rotation rates exceeding 70\% critical.  In the future we hope to address these issues with a more physical treatment of the wind by combining this patch model with a three-dimensional wind radiative transfer code like those presented in \citet{Hennicker2018, Hennicker2019}.

\section{Summary and future work}
In this paper, we have presented the inner workings of the \spamms code as well as several example application cases.  We have shown that the patch model is not only able to accurately reproduce line profiles in spherical cases but is able to reproduce several observed nonspherical effects, such as the Rossiter-Mclaughlin effect and Struve-Sahade effect.  We have also demonstrated the code's ability to fit observed spectra of an overcontact system.  Additionally, we have shown that when taking into account rotational deformations in single stars, the observed temperature and abundances can appear different from spherical models with the same parameters.  Finally, we discussed the advantages of \spamms compared to an alternative code designed for the same purpose, and discussed the limitations of each.

In the future, we plan to improve and apply the \spamms code in several ways.  First, we plan to expand the \fw grid to include other metallicity regimes.  Additionally, we would like to incorporate atmosphere models for other mass regimes so that the code can be used for lower mass systems as well.  While several applications of the code were briefly presented here, we plan to pursue these in further detail in the future.  We plan to analyze all of the massive overcontact binaries currently known and investigate how the results from \spamms compares to the current spherical analysis methods.  Similarly, we plan to systematically investigate how rotational deformation and inclination in single stars affect various parameters such as temperature and abundance when compared to spherical analysis techniques.

\begin{acknowledgements}
We acknowledge support from the FWO-Odysseus program under project G0F8H6N.  Additionally, we acknowledge AS for his thoughtful discussion and helpful comments.
\end{acknowledgements}

\bibliographystyle{aa}
\bibliography{SPAMMS}

\end{document}